\documentclass[12pt]{article}

\usepackage{latexsym}
\usepackage{amsfonts}
\usepackage{amssymb}
\usepackage{epsfig}
\def\N{\mathbb{N}}
\begin{document}

\title{From synchronization to multistability in two coupled quadratic maps}
\author{R. Carvalho\thanks{%
Laborat\'{o}rio de Mecatr\'{o}nica, DEEC, Instituto Superior T\'{e}cnico,
Av. Rovisco Pais, 1096 Lisboa Codex, Portugal}, B. Fernandez\thanks{%
Grupo de F\'{i}sica Matem\'{a}tica, Complexo Interdisciplinar, Universidade
de Lisboa, Av. Gama Pinto 2, 1699 Lisboa Codex, Portugal} \thanks{%
On leave from Centre de Physique Th\'{e}orique, CNRS, Luminy, Marseille,
France} and R. Vilela Mendes\footnotemark[2] }
\date{}
\maketitle

\begin{abstract}
The phenomenology of a system of two coupled quadratic maps is studied 
both analytically and numerically. Conditions for synchronization are given
and the bifurcations of periodic orbits from this regime are identified. In
addition, we show that an arbitrarily large number of distinct stable
periodic orbits may be obtained when the maps parameter is at the
Feigenbaum period-doubling accumulation point. An estimate is given for the
coupling strength needed to obtain any given number of stable orbits.
\end{abstract}

\section{Introduction}

A multistable system is one that possesses a large number of coexisting
attractors for a fixed set of parameters. There is ample evidence for such
phenomena in the natural sciences, with examples coming from neurosciences
and neural dynamics \cite{Schiff} - \cite{Bohr}, optics \cite{Hammel} \cite
{Brambilla}, chemistry \cite{Marmillot} \cite{Laplante} \cite{Hunt},
condensed matter \cite{Prengel} and geophysics \cite{Yoden}. Multistability
also seems to be an essential complexity-generating mechanism in a large
class of agent-based models \cite{Vilela4}.

In view of this, it is important to identify the dynamical mechanisms
leading to multistability and, in particular, to construct simple models
where this phenomenon might be under control. The first mathematical result
in this direction was obtained by Newhouse \cite{Newhouse1} \cite{Newhouse2} 
\cite{Newhouse3} who proved that, near a homoclinic tangency, a class of
diffeomorphisms in a two-dimensional manifold has infinitely many attracting
periodic orbits (sinks), a result that was later extended to higher
dimensions \cite{Palis}. It has also been proved \cite{Colli} that, in
addition to infinitely many sinks, infinitely many strange attractors exist
near the homoclinic tangencies. The stability of the phenomena under small
random perturbations has been studied \cite{Araujo1} \cite{Araujo2}.

A second dynamical mechanism leading to multistability is the addition of
small dissipative perturbations to conservative systems. Conservative
systems have a large number of coexisting invariant sets, namely periodic
orbits, invariant tori and cantori. By adding a small amount of dissipation
to a conservative system one finds that some of the invariant sets become
attractors. Not all invariant sets of the conservative system will survive
when the dissipation is added. However, for sufficiently small dissipation,
many attractors (mainly periodic orbits) have been observed in typical
systems \cite{Poon} \cite{Feudel1} \cite{Feudel2}. The problem of migration
between attractors and their stability in multiple-attractor systems has
also been studied by other authors \cite{Weigel} \cite{Kaneko}. Most of
results are based on numerical evidence. However, using the techniques of
deformation stability \cite{Vilela1} \cite{Vilela2} \cite{Lima} \cite
{Vilela3} some rigorous mathematical results \cite{Vilela5} may be obtained.

Finally, it has been found recently \cite{Carvalho} that, for parameter
values near the Feigenbaum period-doubling accumulation point, quadratic
maps coupled by convex coupling may have a large number of stable periodic
orbits. This is one of the phenomena we study in detail in this paper. The
emphasis on quadratic maps near the Feigenbaum accumulation point has a
motivation close to the idea of control of chaos \cite{Ott} \cite{Control}.
The typical situation in control of chaos, is that of a strange attractor
with an infinite number of embedded periodic orbits, all of them unstable.
These orbits are then stabilized by several methods. If, instead of a large
number of unstable periodic orbits, one has, for example, a large number of
sinks, the controlling situation would seem more promising and robust,
because the control need not be so accurate. It would suffice to keep the
system inside the desired basin of attraction. At the period-doubling
accumulation point the Feigenbaum attractor, because of the properties of
the flip bifurcations, coexists with an infinite set of unstable periodic
orbits. By coupling, as we will show, an arbitrarily large number of orbits
may become stable.

The existence of a large number of stable periodic orbits for just two
coupled quadratic maps, provides a simple model where multistability is well
under control, in the sense that not only the nature of the phenomenon is
completely understood as one may also compute the range of parameters that
provides any desired number of stable orbits. This should be contrasted, for
example, with concrete models for the Newhouse phenomenon \cite{Gambaudo}.

Rather than merely focusing on multistability, we also study the
phenomenology of two coupled quadratic maps, in particular the bifurcations
of periodic orbits and the regime of synchronization.. The stabilization of
orbits in the coupled system is similar to that obtained in higher
dimensional coupled map lattices \cite{Gade} with the exception that, due to
the restricted dimension of the phase space, the types of bifurcations are
different in our system. The results concerning the multistability
phenomenon at $\mu =\mu _{\infty }$ also considerably extend, and also
correct, some imprecise statements in \cite{Carvalho}.

\section{Coupled quadratic maps}

Coupled map lattices (CML) are discrete dynamical systems generated by the
composition of a local nonlinearity and a coupling. The phase space of the
CML considered in this letter is the square $[-1,1]^{2}$ and the dynamics is
generated by the map $F_{\epsilon }$ defined as follows. Given a point $%
(x,y)\in \mathcal{M}$, its image by $F_{\epsilon }$, denoted $(\overline{x},%
\overline{y})$ is given by 
\begin{equation}
\left\{ 
\begin{array}{l}
\bar{x}=(1-\epsilon )f(x)+\epsilon f(y) \\ 
\bar{y}=(1-\epsilon )f(y)+\epsilon f(x)
\end{array}
\right.  \label{DEF}
\end{equation}
where $0\leq \epsilon \leq \frac{1}{2}$ , $f(x)=1-\mu x^{2}$ and $0<\mu < 2$.

The map $f$ maps $[-1,1]$ into itself. Therefore, the convex combination in (%
\ref{DEF}) ensures that $F_{\epsilon }([-1,1]^{2})\subset [-1,1]^{2}$ and
the dynamics is well-defined. We denote the orbit issued from the initial
condition $(x,y)$ by the sequence $\{(x^{t},y^{t})\}_{t\in \N}$, that is to
say, $(x^{0},y^{0})=(x,y)$ and $(x^{t+1},y^{t+1})=F_{\epsilon }(x^{t},y^{t})$
for all $t\in \N$.

For the sake of simplicity, we will often employ the variables $s=\frac{x+y}{%
2}$ and $d=\frac{x-y}{2}$. The previous notation of orbits also applies to
these variables for which relation (\ref{DEF}) becomes 
\begin{equation}
\left\{ 
\begin{array}{l}
\bar{s}=1-\mu ^{2}(s^{2}+d^{2}) \\ 
\bar{d}=-\alpha sd
\end{array}
\right.  \label{DEF2}
\end{equation}
where $\alpha =2\mu (1-2\epsilon )$.

Finally, note that the dynamics commutes with the symmetry $%
\sigma(s,d)=(s,-d)$ or $\sigma(x,y)=(y,x)$ in the original variables.

\section{Synchronization}

If $d^{t}=0$, then $d^{t+1}=0$ and $s^{t+1}=f(s^{t})$. In this case, the
orbit is said to be synchronized (from $t$ on). More generally, an orbit is
said to synchronize if $\lim_{t\rightarrow \infty }d^{t}=0$ and if all
orbits synchronize, then we say to have synchronization of the map.
Synchronization is the simplest dynamical regime exhibited by
two-dimensional CML.

To determine a sufficient condition for synchronization in our system, we
note that for any orbit, one has $|s^{t}|\leq 1$ for all $t\in \N$. It
follows from (\ref{DEF2}) that the condition $|\alpha |<1$ ensures an
exponential decay of $|d^{t}|$, and hence synchronization.. Since $\alpha
\geq 0$, the condition $|\alpha |<1$ is equivalent to the following ones
(see Figure 1). 
\[
\epsilon =\frac{1}{2}\quad \textnormal{or}\quad \epsilon <\frac{1}{2}\quad \textnormal{%
and}\quad \mu <\mu _{1}(\epsilon ):=\frac{1}{2(1-2\epsilon )}.
\]
\begin{figure}[tbh]
\begin{center}
\psfig{figure=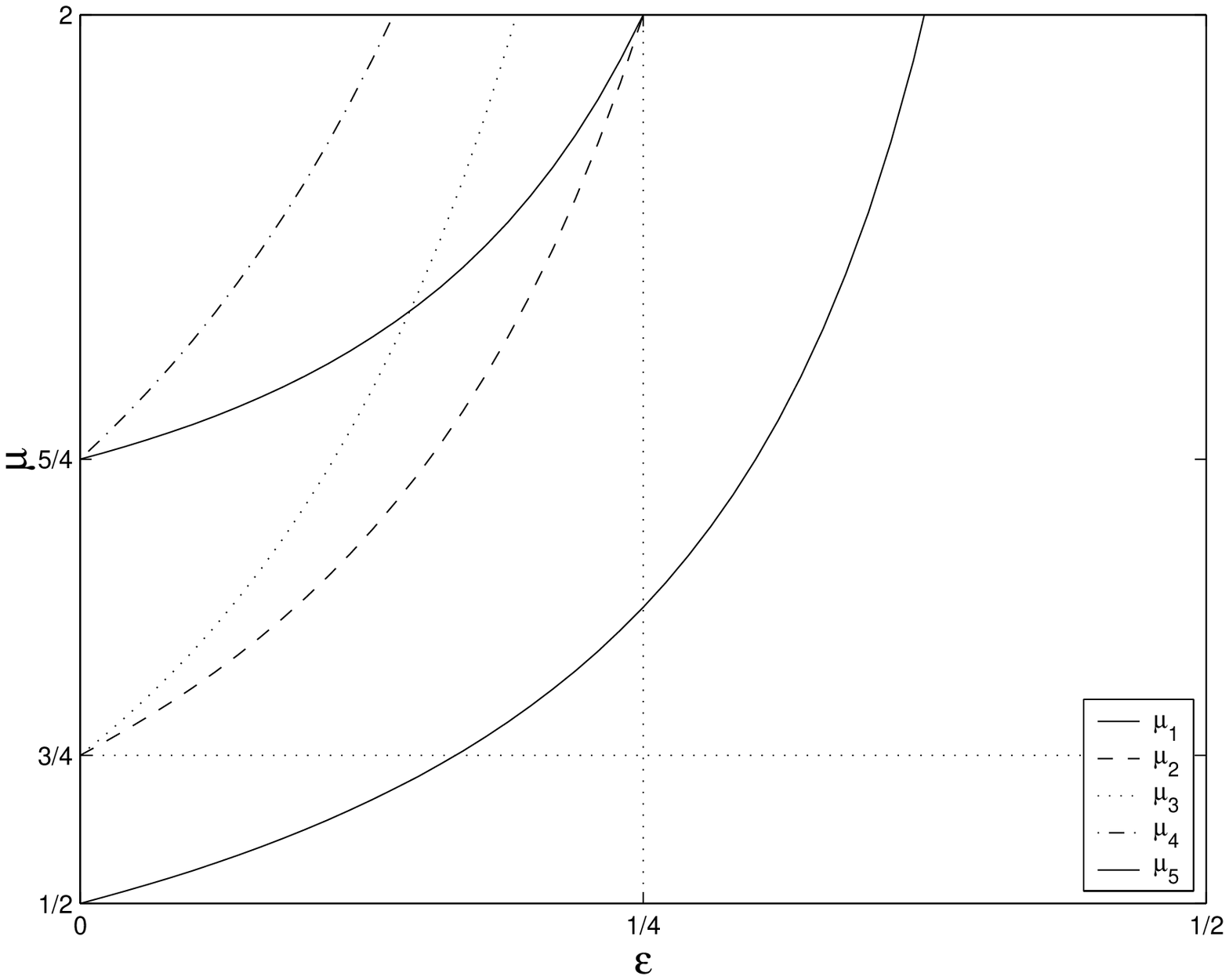,width=12truecm}
\end{center}
\caption[]{Synchronization curve $\mu =\mu _{1}(\epsilon )$ and other
bifurcations curves. $\mu _{2}(\epsilon )$ is the period-doubling of the
synchronized fixed point (i.e.\ birth of the phase opposition period-2
orbit), $\mu _{3}(\epsilon )$ is the pitchfork of the phase opposition
period-2 orbit and $\mu _{4}(\epsilon )$ the Hopf bifurcation of the same
orbit. $\mu _{(4)}(\epsilon )$ is the period-doubling of the synchronized
period-2 orbit (i.e.\ birth of the phase opposition period-4 orbit)}
\end{figure}
From now on, we assume that $\epsilon <\frac{1}{2}$. The condition $\mu <\mu
_{1}(\epsilon )$ is not necessary. Indeed, if for instance $\mu <\frac{3}{4}$%
, then $f$ has an attracting fixed point in $[-1,1]$, and one can prove that
synchronization occurs. When $\epsilon $ is sufficiently small, this happens
even though $\mu \geq \mu _{1}(\epsilon )$.

\section{Non-synchronized period-2 orbits}

Starting from synchronization and modifying the parameters, non-synchronized
(periodic) orbits appear from bifurcations of synchronized (periodic) ones.
To understand this phenomenon, as well as the bifurcations of subsequent
orbits, we now study analytically the periodic orbits of period 1 and 2. Let 
$(s^{*},0)$ be the synchronized fixed point denoted in $(s,d)$-variables. It
exists for any values of the parameters and $F_{\epsilon }$ has no other
fixed point in $[-1,1]^{2}$. In $(s,d)$-variables, the Jacobian of $%
F_{\epsilon }$ at this fixed point is diagonal. One eigenvalue is $f^{\prime
}(s^{*})$ and the corresponding eigendirection is the diagonal $d=0$. The
other eigenvalue is $(1-2\epsilon )f^{\prime }(s^{*})$ and the corresponding
direction, orthogonal to the diagonal is referred as the anti-diagonal.

The condition $f^{\prime }(s^{*})=-1$ then determines a period-doubling
bifurcation, which is of co-dimension 1 if $\epsilon >0$. This is the
well-known period-doubling bifurcation of $f$ which creates a synchronized
period-2 orbit of $F_{\epsilon }$.

Moreover, one checks that the derivative $f^{\prime }(s^{*})$ is negative
for any $\mu $. Hence the conditions $\epsilon >0$ and $(1-2\epsilon
)f^{\prime }(s^{*})=-1$, i.e.\ 
\[
\epsilon >0\quad \textnormal{and}\quad \mu =\mu _{2}(\epsilon ):=\frac{3-4\epsilon 
}{4(1-2\epsilon )^{2}},
\]
determine another co-dimension 1 period-doubling bifurcation of the
synchronized fixed point. Indeed the conditions of the corresponding
bifurcation theorem (see e.g.\ \cite{Guckenheimer}) are satisfied when the
curve $\mu =\mu _{2}(\epsilon )$ is crossed upward.

The period-2 orbit created at this bifurcation is non-synchronized and
symmetric. To show this, denote by $(s_{1},d_{1})$ and $(s_{2},d_{2})$ its
components. Since the multiplier $(1-2\epsilon )f^{\prime }(s^{*})$ is
negative and the bifurcating direction is the anti-diagonal, we have $%
d_{1}d_{2}<0$ (sufficiently close to the bifurcation). Because of the $%
\sigma $ symmetry, the map $F_{\epsilon }$ also has a period-2 orbit with
components $(s_{1},-d_{1})$ and $(s_{2},-d_{2})$. Consequently if $\sigma
(s_{1},d_{1})\neq (s_{2},d_{2})$, the system would have two periodic orbits
created by a co-dimension 1 bifurcation. This is impossible by the unicity
in the bifurcation theorem. Therefore, sufficiently close to the
bifurcation, we have $\sigma (s_{1},d_{1})=(s_{2},d_{2})$ and $d_{1}\neq 0$
which is the desired conclusion.

By continuity in the parameters of $F_{\epsilon }$, sufficiently close to
the bifurcation, this symmetric orbit is stable with respect to
perturbations in one direction (the anti-diagonal direction at the
bifurcation) and since $f^{\prime }(s^{*})<(1-2\epsilon )f^{\prime
}(s^{*})<-1$, it is unstable in the direction orthogonal to the latter.

The bifurcations will now be computed. The orbit with $\sigma
(s_{1},d_{1})=(s_{2},d_{2})$ and $d_{1}\neq 0$ exists for any $\mu >\mu
_{2}(\epsilon )$ and is the unique (up to time translations) period-2
non-synchronized symmetric orbit of $F_{\epsilon }$ in $[-1,1]^{2}$.

Computing the corresponding Jacobian, one obtains the equation for the
multipliers 
\[
\lambda ^{2}-2\left[ \left( 1-2\varepsilon \right)
(S^{2}-D^{2})+2\varepsilon ^{2}S^{2}\right] \lambda +\left( 1-2\varepsilon
\right) ^{2}(S^{2}-D^{2})^{2}=0 
\]
where 
\[
\left\{ 
\begin{array}{l}
S=2\mu s_{1}=\frac{1}{1-2\varepsilon } \\ 
D=2\mu d_{1}=\frac{\sqrt{4\mu (1-2\varepsilon )^{2}-(3-4\varepsilon )}}{%
1-2\varepsilon }
\end{array}
\right. 
\]
Direct calculations show that, if $\epsilon>0$, the multipliers, say $%
\lambda _{-}$ and $\lambda _{+}$, have zero imaginary part iff $\mu \leq 
\frac{(2-3\varepsilon )^{2}}{4(1-2\varepsilon )^{3}}$. Under this condition,
we have $0<\lambda _{-}<1$ if $\mu>\mu_2(\epsilon)$ and $\lambda_+>1$ iff $%
\mu_2(\epsilon)<\mu<\mu_3(\epsilon):=\frac{3}{4(1-2\epsilon)^2}$. (The
inequality $\mu_2(\epsilon)<\mu_3(\epsilon)$ indeed holds if $0<\mu<2$ and $%
0<\epsilon<\frac{1}{2}$, see Figure 1.)

Consequently, by increasing $\mu $, the symmetric orbit suffers an inverse
pitchfork bifurcation at $\mu =\mu _{3}(\varepsilon )$. This bifurcation is
generic for a symmetric orbit in a system with symmetry \cite{Guckenheimer}
and the conditions of the bifurcation theorem hold when the curve $\mu =\mu
_{3}(\varepsilon )$ is crossed upward.

This bifurcation creates two non-symmetric period-2 orbits (one orbit and
its symmetric). We have checked that these orbits exist for any $\epsilon
\geq 0$ and $\mu >\mu _{3}(\epsilon )$. For $\epsilon =0$, their components
are combinations of a fixed point of $f$ and the components of a period-2
orbit.

When the imaginary part of $\lambda_-$ and $\lambda_+$ is not zero, we have $%
|\lambda_- |=|\lambda_+ |<1$ iff $\mu<\mu_4(\epsilon):=\frac{5-6\epsilon}{%
4(1-2\epsilon)^2}$. (Once again, if $0<\mu<2$ and $0<\epsilon<\frac{1}{2}$,
the inequality $\mu_3(\epsilon)<\mu_4(\epsilon)$ is satisfied, see Figure 1.)

The symmetric orbit is thus stable in the interval $\mu _{2}(\epsilon )<\mu
<\mu _{4}(\epsilon )$. If $\epsilon >0$ and the curve $\mu =\mu
_{4}(\epsilon )$ is crossed upward, this orbit suffers a Hopf bifurcation
creating a locally stable invariant circle. A numerical calculation shows
that the latter is destroyed when $\mu $ is sufficiently large or when $%
\epsilon $ is sufficiently small. Obviously, if $\epsilon =0$, it does not
exist and the bifurcation at $\mu =\mu _{4}(0)=\frac{5}{4}$ which is a
period-doubling bifurcation of $f$ creates a period-4 orbit.

Note that invariant circles in two-dimensional CML resulting from the
destabilization of a symmetric orbit and their normal form had already been
reported in \cite{Kurten}. In that work, the system is also defined by (\ref
{DEF}), but the local map is $f(x)=ax(1-x)$ and $\epsilon $ may be larger
than $\frac{1}{2}$. 
\begin{figure}[tbh]
\begin{center}
\psfig{figure=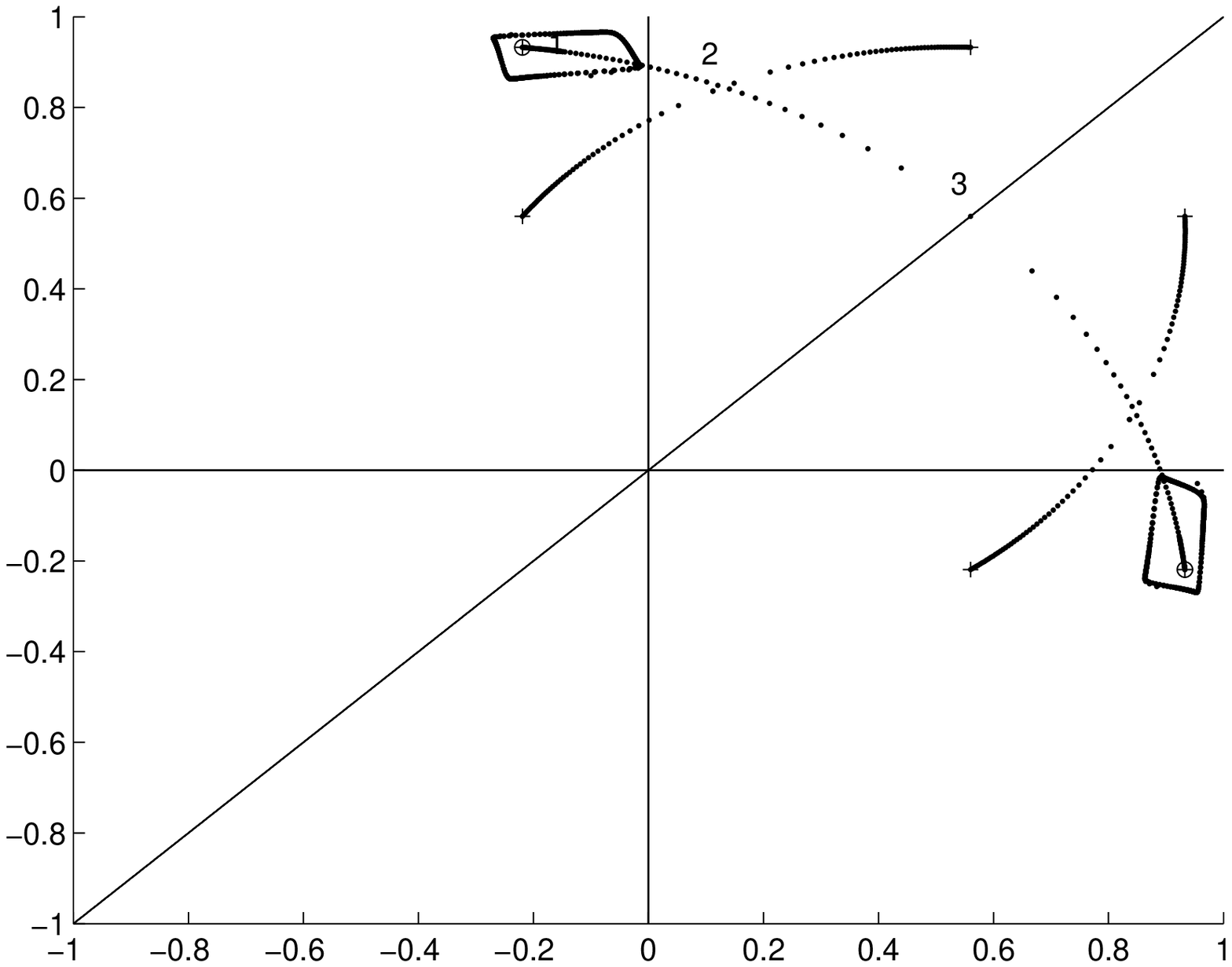,width=12truecm}
\end{center}
\caption[]{$\varepsilon -$evolution of a period-2 orbit for $\mu =\mu _{\infty
} $. An example of invariant circle is also shown for $\epsilon =0.03$}
\end{figure}
Figure 2 shows an example of the phenomenology described above. Numerically,
it is more convenient to follow the orbits from $\epsilon =0$ to increasing
values of the coupling. In this picture, as well as in the following ones
the map parameter is $\mu =\mu _{\infty }$ (the accumulation point of the
period-doubling cascade). In Figure 2, from $\epsilon =0$ (the circle) to $%
\epsilon =0.039$ (the point labelled 1), the symmetric orbit is unstable.
The figure also shows the invariant circle for $\epsilon =0.03$. Between the
points 1 and 2, the symmetric orbit is stable. At point 2, the pitchfork
occurs, the symmetric orbit becomes unstable and the non-symmetric orbits
are created. Finally, the point 3 ($\epsilon =0.182$) corresponds to the
collapse on the synchronized fixed point.

\section{The phase opposition orbits}

The previous phenomenology is not restricted to small periods but extend to
any power of 2. In particular, the synchronized period-$2^{p}$ orbit may
destabilize to create a symmetric (non-synchronized) orbit of twice the
period.

Given $p\in \N$, let $\{s_{i}\}_{1\leq i\leq 2^{p}}$ be the components of
the period-$2^{p}$ orbit of $f$. The points $\{(s_{i},0)\}_{1\leq i\leq
2^{p}}$ are the components of the synchronized period-$2^{p}$ orbit of $%
F_{\epsilon }$. By the chain rule and since each Jacobian at $(s_{i},0)$ is
diagonal, the corresponding multiplier along the anti-diagonal direction is 
\[
(1-2\epsilon )^{2^{p}}\prod_{i=1}^{2^{p}}f^{\prime }(s_{i})
\]
The condition that this multiplier equals $-1$ determines, if $\epsilon >0$,
a co-dimension 1 period-doubling bifurcation. Applying the reasoning of the
previous section to each component $(s_{i},0)$, we conclude that this
bifurcation creates an orbit with the property $\sigma
(s^{t+2^{p}},d^{t+2^{p}})=(s^{t},d^{t})$ and $d^{t}\neq 0$ for all $t\in \N$%
, which is called a phase-opposition period-$2^{p+1}$ orbit.

Since $(1-2\epsilon )^{2^{p}}<1$, this bifurcation occurs only if the
bifurcation along the diagonal direction has occurred (the local
period-doubling bifurcation of $\{s_{i}\}_{1\leq i\leq 2^{p}}$). In other
words, the phase opposition period-$2^{p}$ orbit exists only if the
synchronized period-$2^{p}$ orbit does. 
\begin{figure}[tbh]
\begin{center}
\psfig{figure=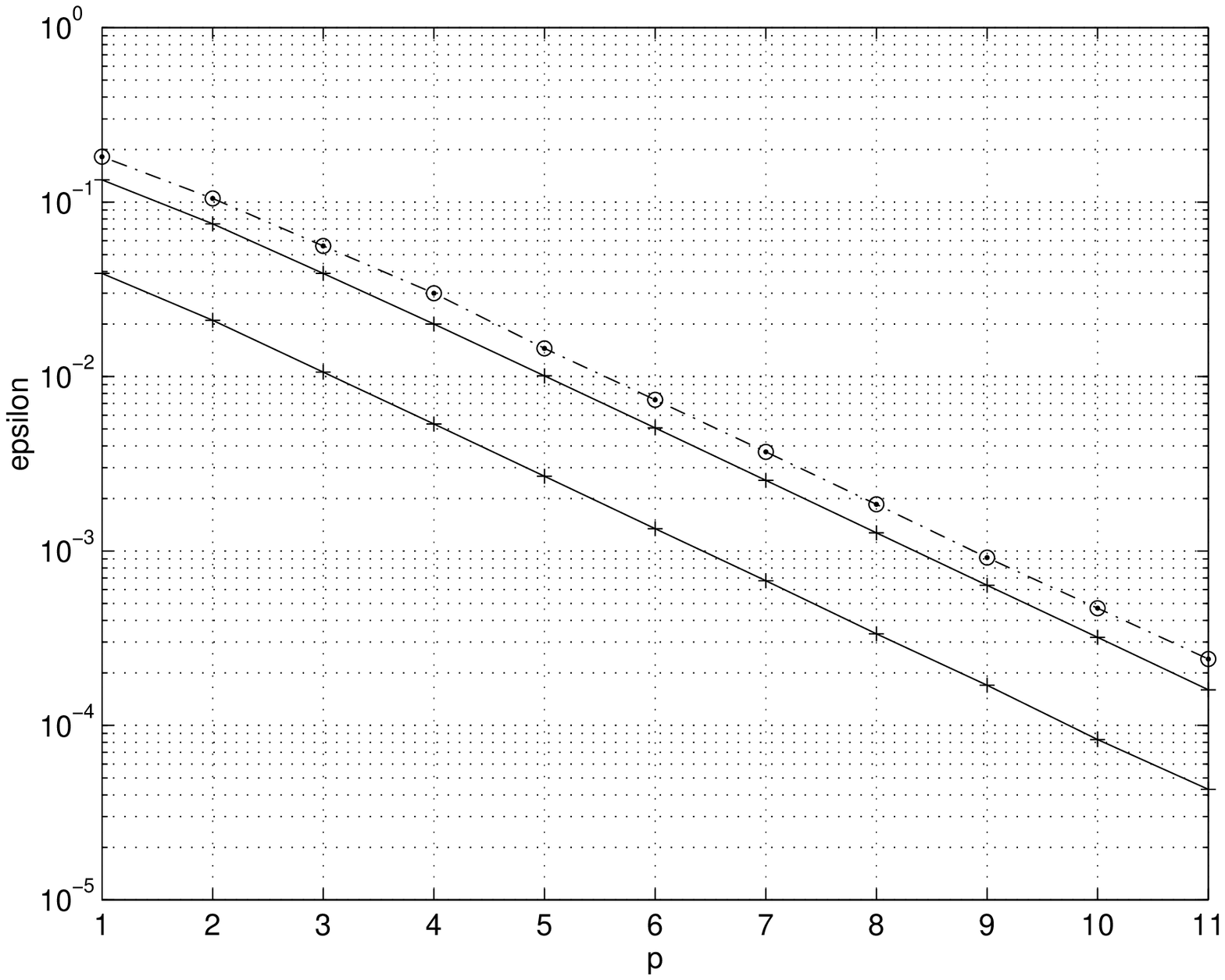,width=12truecm}
\end{center}
\caption[]{Bifurcation values of $\epsilon $ for the phase opposition orbits.
From top to bottom, birth, pitchfork and Hopf bifurcation}
\end{figure}
Moreover it follows from Figure 3 that, at least for $\mu =\mu _{\infty }$,
the phase opposition period-$2^{p+1}$ orbit exists only if the phase
opposition period-$2^{p}$ orbit does. This is confirmed analytically for the
period$-4$ orbit whose existence condition is the instability of the
synchronized period$-2$ orbit in the anti-diagonal direction. One obtains 
\[
\mu >\mu _{(4)}(\epsilon )=1+\frac{1}{4(1-2\epsilon )^{2}},
\]
and $\mu _{2}(\epsilon )<\mu _{(4)}(\epsilon )$ if $0<\epsilon <\frac{1}{4}$
and $\mu _{2}(\epsilon )\geq 2$ and $\mu _{(4)}(\epsilon )\geq 2$ if $\frac{1%
}{4}\leq \epsilon <\frac{1}{2}$ (see Figure 1).

Furthermore, a numerical calculation at $\mu =\mu _{\infty }$ , reported in
Figure 3, shows that the succession of bifurcations of a phase opposition
orbit does not depend on the period. On this picture, we have plotted the
values of $\epsilon $ for the Hopf bifurcation, the pitchfork bifurcation
and the period-doubling bifurcation creating the orbit, versus the power of
the period. For each period, the phenomenology is identical to that
described in the previous section, with an adequate change of scale in $%
\epsilon $. In addition, the picture shows that several phase opposition
orbits may be stable for $\epsilon >0$ fixed. This stabilization is an
effect of the coupling that will be discussed below.

Finally, since the phase opposition orbits are the first orbits to appear
when the parameters are varied from synchronization and since the first such
orbit that is created is of period 2, it follows that a necessary and
sufficient condition for synchronization is $\mu \leq \mu _{2}(\epsilon )$,
the condition for the existence of the latter. 
\begin{figure}[tbh]
\begin{center}
\psfig{figure=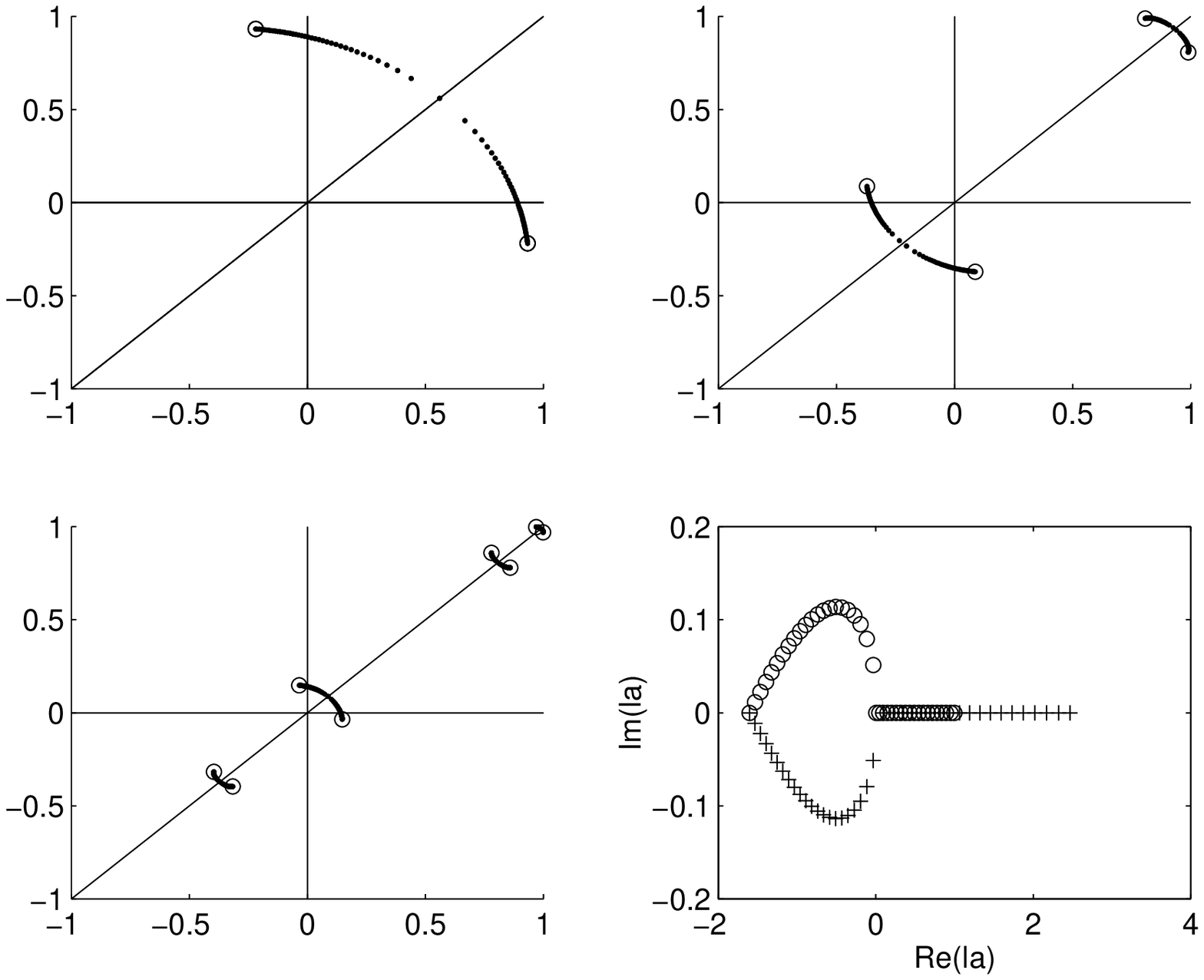,width=12truecm}
\end{center}
\caption[]{$\varepsilon -$evolution of distance$-1$ orbits and their
multipliers at $\mu =\mu _{\infty }$}
\end{figure}

\section{The non-symmetric orbits}

We now analyze the existence and the stability of other period-$2^{p}$
orbits for $\mu =\mu _{\infty }$. Our interest for this value of $\mu $ is
that the scaling properties of $f$ are reflected on scaling laws for the
periods and values of $\epsilon $ at which the bifurcations occur (see
Figure 3 and 8). We only consider the orbits which for $\epsilon =0$ have
the same period on projection to both axis $x$ and $y$. These orbits are
followed numerically when $\epsilon $ increases and are referred using the
phase shift of their components at $\epsilon =0$. 
\begin{figure}[tbh]
\begin{center}
\psfig{figure=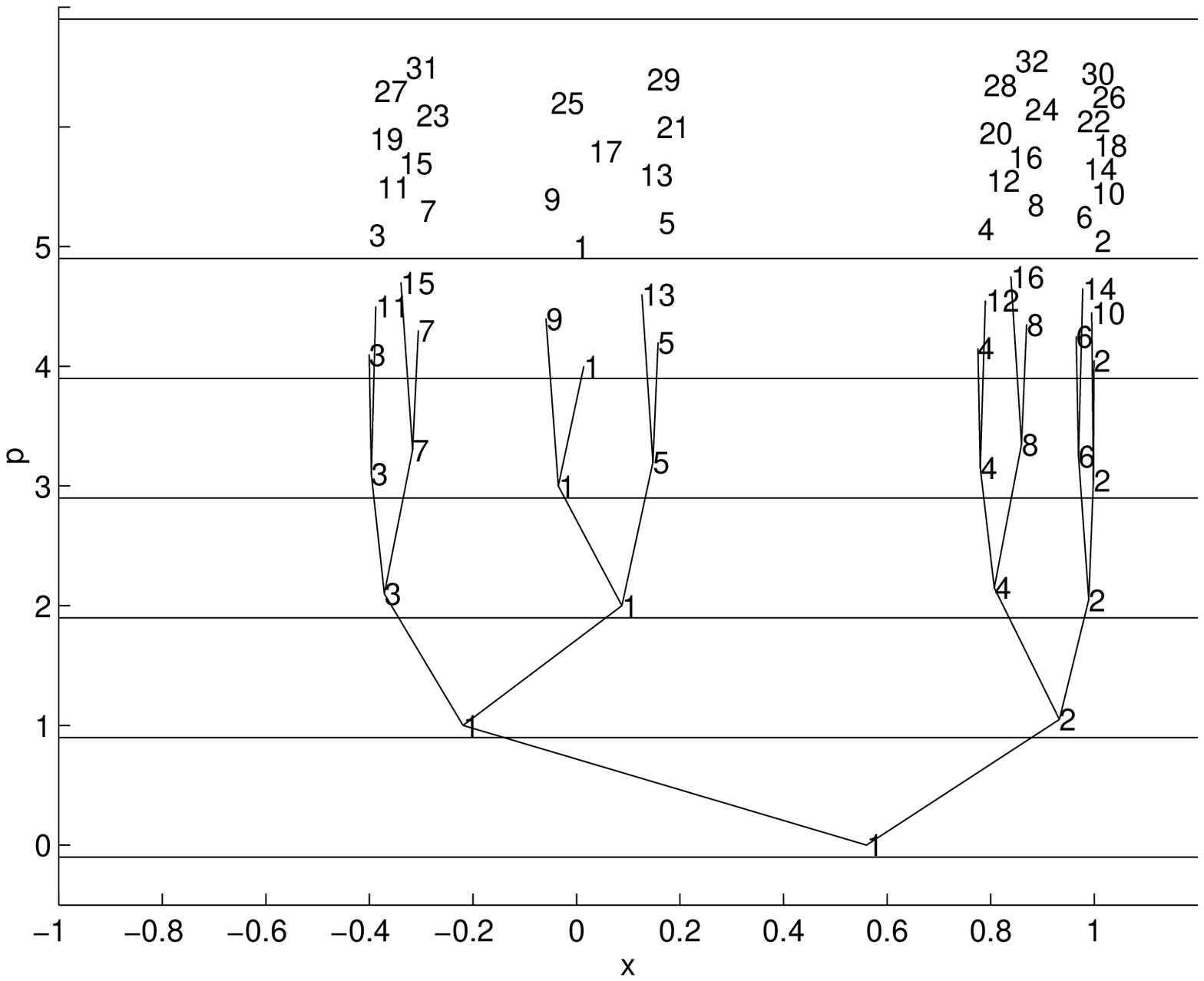,width=12truecm}
\end{center}
\caption[]{Periodic orbits for the one-dimensional map at $\mu =\mu _{\infty }$}
\end{figure}
For $\mu =\mu _{\infty }$, the map $f$ has a period-$2^{p}$ orbit for each $%
p\in \N$, whose components for $p$ up to 5 are shown in Figure 5. In this
picture, the numbers reflect the order in which the components are visited
and the tree structure represents the origin of each component in the
bifurcation cascade. An important notion is the dyadic distance $\delta $
between the components of an orbit. $\delta $ is the number of steps one has
to go back in the bifurcation tree to meet a common component.

The dyadic distance is used to characterize the families of periodic orbits
that we are considering. For instance, the coordinates of each component of
a synchronized orbit are at distance 0, those of a phase-opposition orbit
are at distance 1. Accordingly, when we speak of distance$-k$ orbit we refer
to the dyadic distance of the coordinates of its components.

For any $\delta\geq 1$, there are $2^{\delta -1}$ different orbits with
distance $\delta $ which have coordinates out of phase by $2^{p-\delta
}+\alpha _{1}2^{p-\delta +1}+\alpha _{2}2^{p-\delta +2}+\cdots +\alpha
_{d-1}2^{p-1}$ steps, with $\alpha _{i}\in \left\{ 0,1\right\} $. The
distance of a period-$2^p$ orbit is at most $p$ ($p\in\N$). 
\begin{figure}[htb]
\begin{center}
\psfig{figure=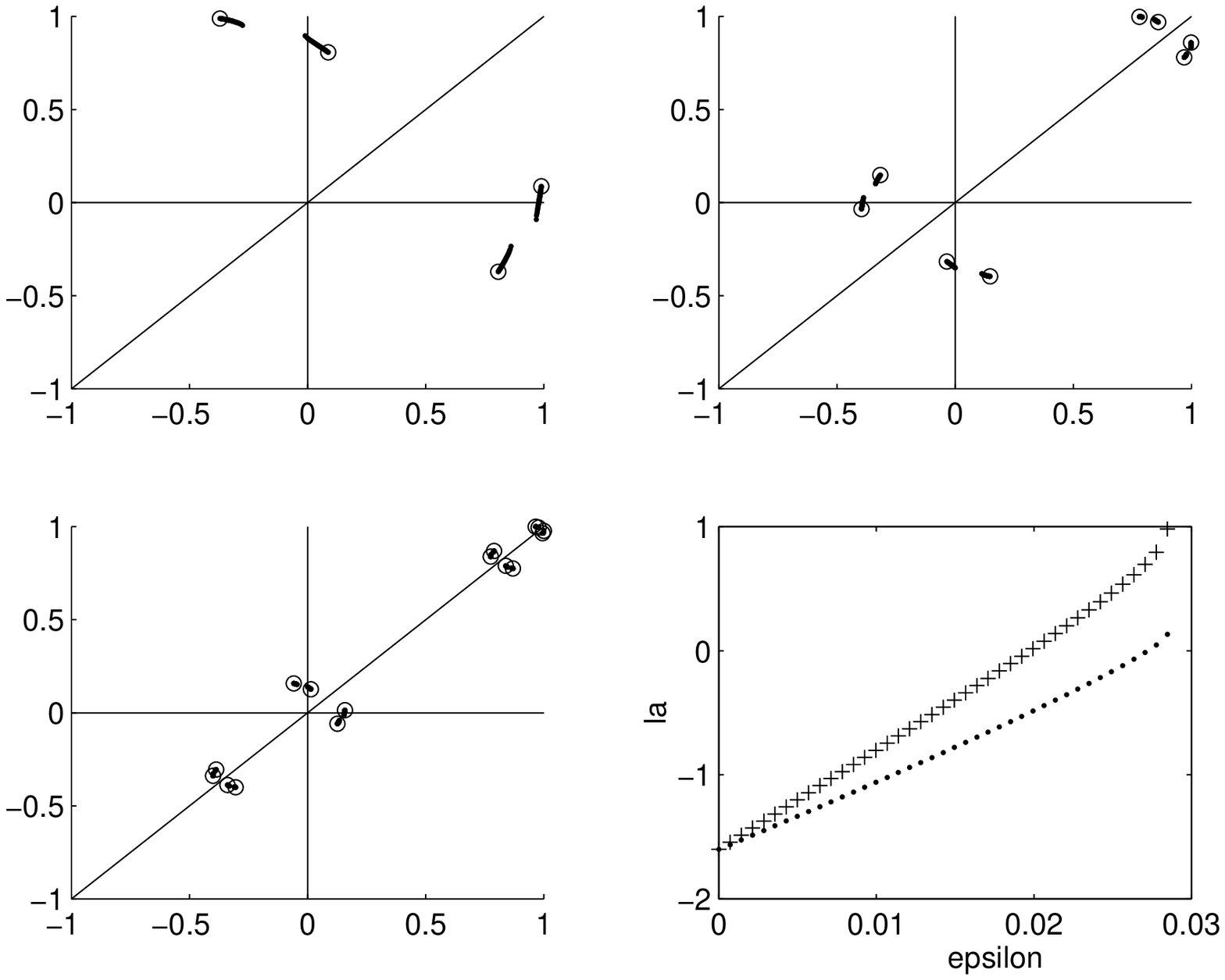,width=12truecm}
\end{center}
\caption[]{$\varepsilon -$evolution of distance$-2$ orbits and their
multipliers at $\mu =\mu _{\infty }$}
\end{figure}
For $\epsilon =0$, the only symmetric orbits are those at distance 0 and 1.
This property is preserved for $\epsilon >0$ as shows Figure 6 for $\delta
=2 $. The succession of bifurcations of orbits with distance $\delta \geq 2$
should then differ from those with distance 1. The differences are seen in
Figure 6 which shows the evolution of the eigenvalues. For $\epsilon =0$,
the orbit is unstable. When $\epsilon $ increases, it suffers two collisions
with orbits of twice the period when the eigenvalues cross $-1$ and then
becomes stable. (When decreasing $\epsilon $, these collisions would be
period-doubling bifurcations.) If $\epsilon $ increases further, the orbit
collides with an unstable one of the same period in a saddle-node
bifurcation when the larger eigenvalue reaches 1. For larger values of $%
\epsilon $, the orbit does not exist. The unstable orbit with which it
collides is the one that at $\epsilon =0$ has period $2^{p}$ in one
projection and $2^{p-1}$ in the other. For higher dyadic distances, the
overall variation of the eigenvalues is similar to the $\delta =2$ case. 
\begin{figure}[htb]
\begin{center}
\psfig{figure=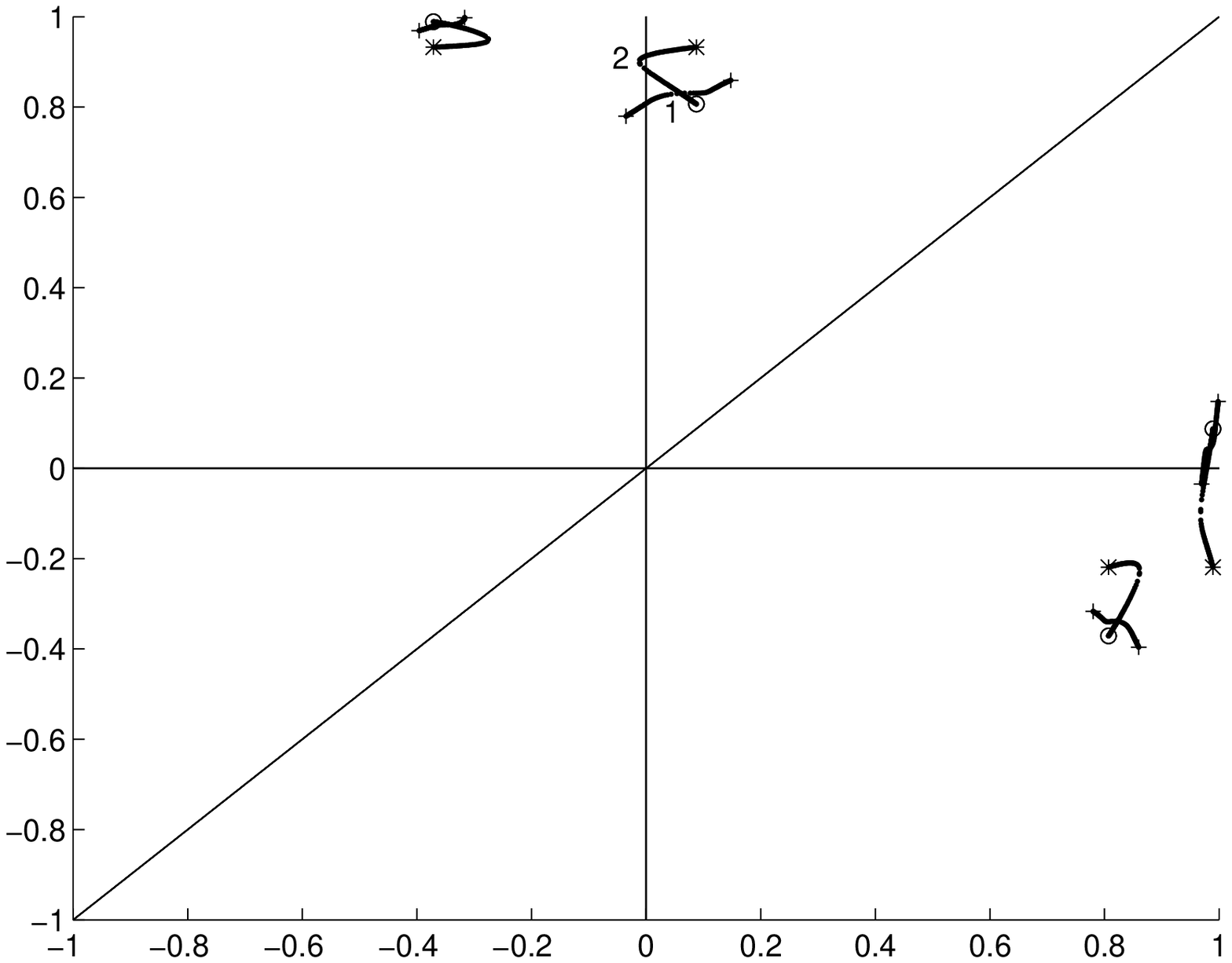,width=12truecm}
\end{center}
\caption[]{$\varepsilon -$evolution of a period$-4$ distance$-2$ orbit and its
bifurcating orbits at $\mu =\mu _{\infty }$}
\end{figure}
Figure 7 shows a typical example of these phenomena for the case $\delta =2$%
. Between $\epsilon =0$ and the point labelled 1 in the figure, the orbit is
unstable. The point 1 corresponds to the smaller eigenvalue crossing -1 (see
Figure 6). Therefore, between the point 1 and 2, the orbit is stable. It
disappear at the point 2 when it collides with an unstable orbit of the same
period.

\section{Multistability}

We have seen that the coupling stabilizes the orbits with distance larger
than 0 at $\mu =\mu _{\infty }$. There are indeed two mechanisms responsible
for this stabilization.. The determinant of the Jacobian of a period-$2^{p}$
orbit is 
\[
(1-2\epsilon )^{2^{p}}(-2\mu _{\infty
})^{2^{p+1}}\prod_{i=1}^{2^{p}}x_{i}y_{i}.
\]
The term $(1-2\epsilon )^{2^{p}}$ coming from the coupling decreases when $p$
increases. However, there is yet a second stabilizing mechanism. Denote by $%
\Gamma (\epsilon )$ the remaining factor in the determinant 
\[
\Gamma (\epsilon )=(-2\mu _{\infty })^{2^{p+1}}\prod_{i=1}^{2^{p}}x_{i}y_{i}.
\]
Without coupling, $\Gamma (0)$ is simply the square of the multiplier of $f$
for the periodic orbit. From the properties of the Feigenbaum - Cvitanovic
functional equation it follows \cite{Carvalho} that this factor converges to
a fixed value around $-1.6$ when $p$ increases. The coupling however,
changes the position of the orbit components in such a way that this factor
also decreases. It is the combined action of this decrease with the
contraction of the coupling that brings the eigenvalues into the interior of
the unit circle and stabilizes the orbits.

For small $\epsilon $ there is a simple geometrical interpretation for the
variation of $\Gamma (\epsilon )$. The reason why in the one dimensional map
the product $\prod_{i=1}^{2^{p}}x_{i}$ remains constant, when $p$ grows, is
because each time the period doubles, the doubling in the number of factors
greater than one is compensated by the fact that the component of the orbit
closest to zero approaches zero a little more. 
\begin{figure}[tbh]
\begin{center}
\psfig{figure=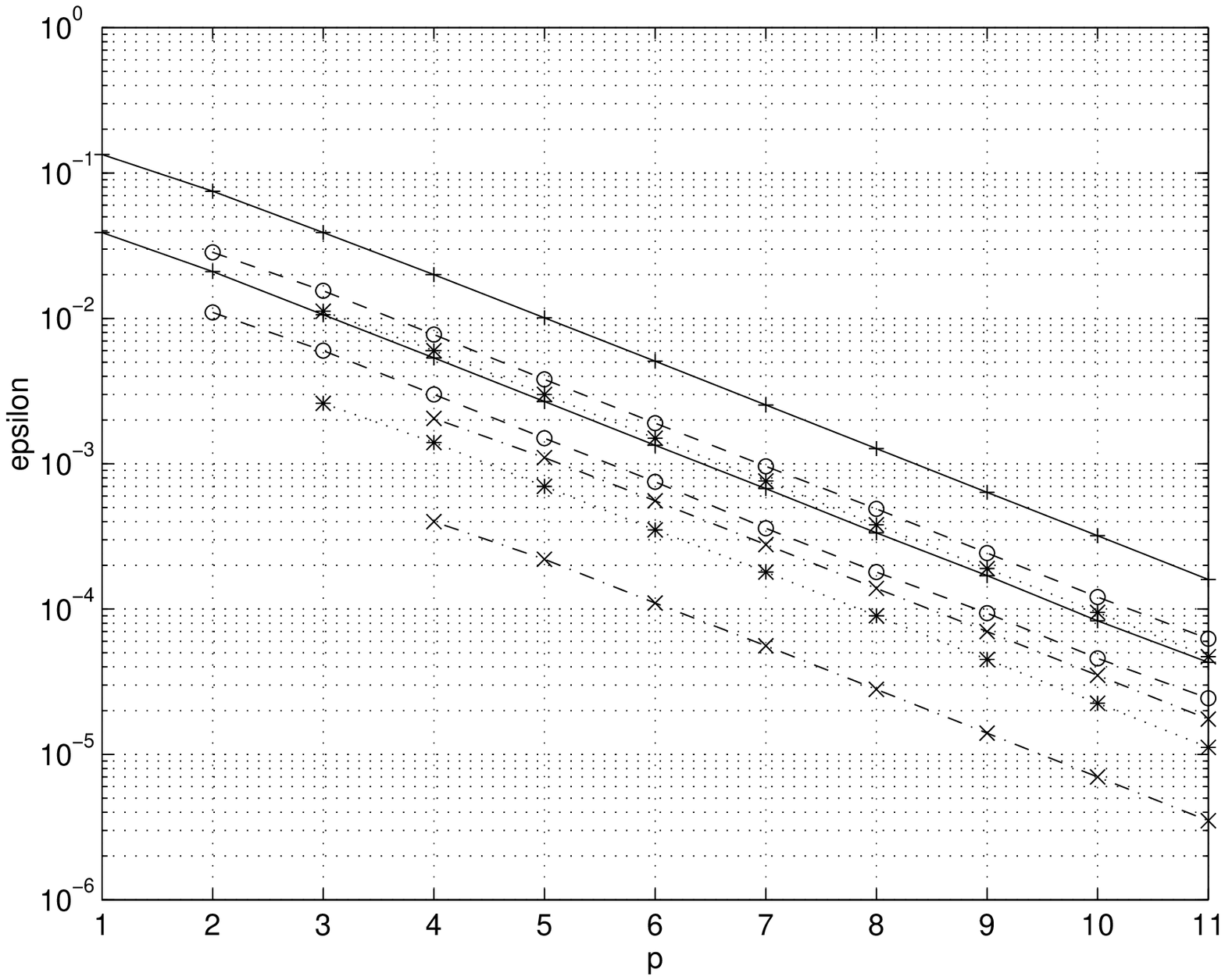,width=12truecm}
\end{center}
\caption[]{Stability lines of orbits from distance $1$ (top) to $4$ (bottom)}
\end{figure}
For the unstable orbits along the period-doubling chain, the orbit
components closest to zero alternate on each side of the origin. The
contracting effect of the convex coupling tends to bring the orbits back in
the period-doubling hierarchy. Therefore, because the component closest to
zero has to move across the origin for the orbit to approach the one with
half the period, this implies that the product of the coordinates is going
to decrease. The greater the dyadic distance between the orbit projections
on the axis, the greater will be the perturbation that the original
(one-dimensional) orbits suffer. Therefore one expects the contracting
effect in $\Gamma (\epsilon )$ to increase with the dyadic distance. This
effect is quite apparent on Figure 8 which shows the stabilizing and
destabilizing lines for orbits with distance from 1 to 4. The shift
downwards of the stable regions for successively larger dyadic distances
implies that the smaller $\epsilon $ is, the larger the number of distinct
stable orbits that are obtained. An accurate numerical estimate of the
number of distinct orbits is obtained by computing the derivative $D\left(
p,\delta \right) =\partial \left( \prod_{i=1}^{2^{p}}x_{i}y_{i}\right)
/\partial \epsilon $ at $\epsilon =0$ for each $p$ and dyadic distance $%
\delta $. Actually this derivative provides an accurate estimate of $\Gamma
\left( \epsilon \right) $ itself, because this one varies almost linearly
with $\epsilon $ for most of the stable range of the orbits. On Figure 9,
the scaling properties, when $p$ grows, of this derivative are shown. From
these results one computes 
\[
\log \left( -D\left( p,\delta \right) \right) =g\left( \delta \right)
-2^{p+1}
\]
with 
\[
g\left( \delta \right) \simeq 0.907\delta +3.475
\]
Notice that in Figure 9 there is more than one data point for each pair $%
\left( p,\delta \right) $ which correspond to non-equivalent orbits with the
same dyadic distance. 
\begin{figure}[tbh]
\begin{center}
\psfig{figure=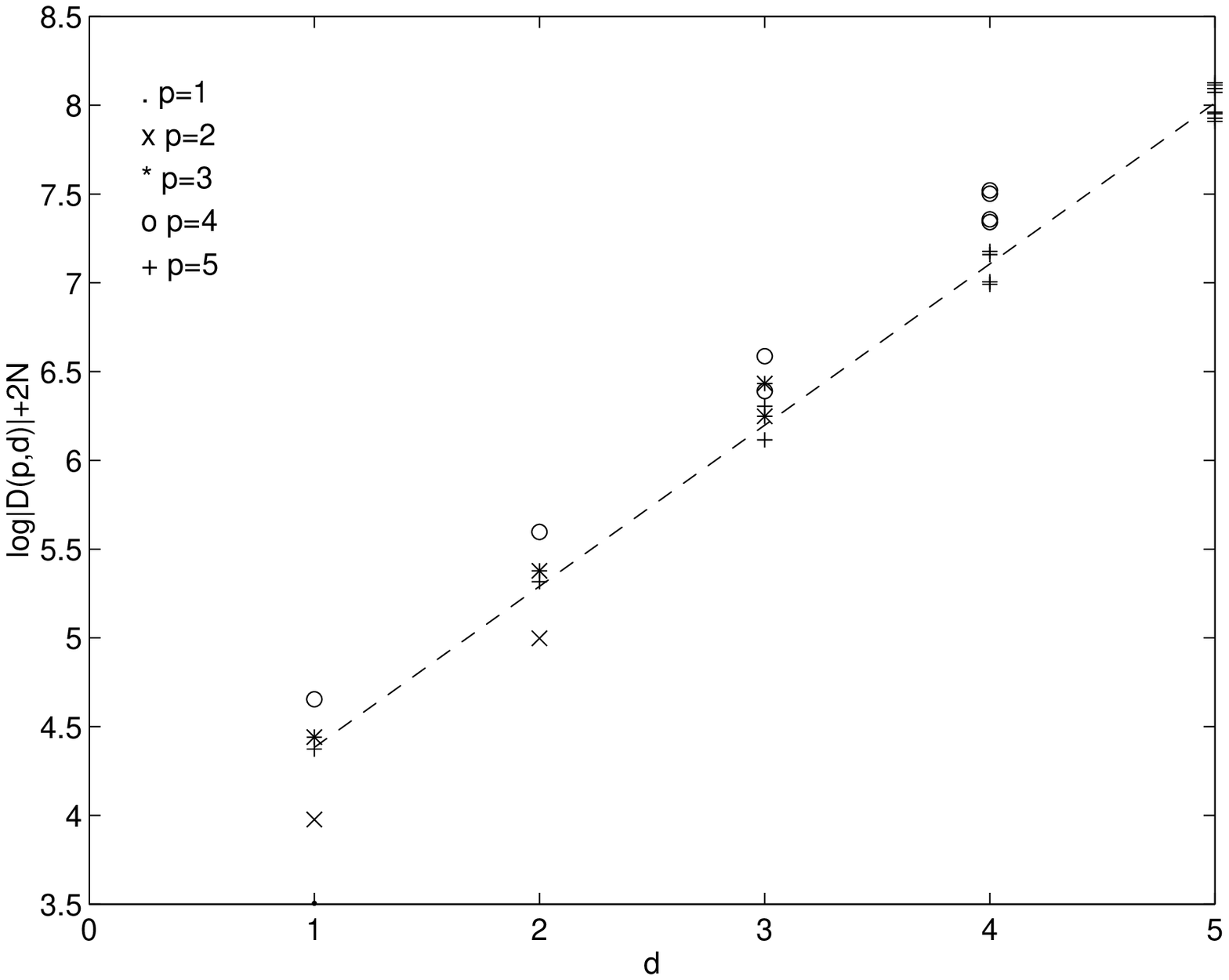,width=12truecm}
\end{center}
\caption[]{Scaling properties of $D\left( p,\delta \right) $}
\end{figure}
Two other useful estimates are:

\begin{itemize}
\item  the value of the smallest $\epsilon $ parameter that stabilizes an
orbit of dyadic distance $\delta $ equal to the power $p$ 
\[
\log \epsilon _{\delta =p}^{\min }=-1.535p-1.547
\]

\item  the value of the largest $\epsilon $ parameter for which a $\delta =1$
orbit is stable 
\[
\log c_{\delta =1}^{\max }=-0.679p-1.235
\]
\end{itemize}

From this, one obtains the result that at least $k$ distinct stable orbits
are obtained if 
\[
0<\epsilon \lesssim \exp \left( -0.99-1.22k\right) 
\]
$k$ is only a lower bound on the number of distinct stable periodic orbits,
because here we have studied only orbits with the same period under
projection in the two axis.

In conclusion: \textit{for sufficiently small }$\epsilon $\textit{\ an
arbitrarily large number of distinct stable periodic orbits is obtained}.
However, for any fixed $\epsilon $, it is an arbitrarily large number that
is obtained, not an infinite number. Most orbits either synchronize (and are
then unstable) or disappear as $\epsilon $ grows. As a result, a reasoning
based on the implicit function theorem, as used in \cite{Carvalho} is
misleading. Given a sequence of orbits of different periods, even if they
remain as orbits for a small perturbation, that does not mean that their
(smallest) periods remain distinct.

\end{document}